\documentclass[12pt,a4,portrait]{article}
\usepackage{graphicx}
\usepackage{latexsym}
\usepackage{hyperref}

\newcommand{\be}{\begin{equation}}
\newcommand{\ee}{\end{equation}}
\newcommand{\beq}{\begin{eqnarray}}
\newcommand{\eeq}{\end{eqnarray}}
\newcommand{\bed}{\begin{displaymath}}
\newcommand{\eed}{\end{displaymath}}
\newcommand{\bs}{\begin{slide}}
\newcommand{\es}{\end{slide}}
\newcommand{\bc}{\begin{center}}
\newcommand{\ec}{\end{center}}
\newcommand{\bi}{\begin{itemize}}
\newcommand{\ei}{\end{itemize}}
\begin{document}
\thispagestyle{empty}

{\phantom .}

\begin{center}

\vspace{3cm}

{\Large\bf{Scalar-tensor cosmology at the general relativity limit: Jordan
vs Einstein frame}}

\vspace{1cm}

Laur J\"arv,$^{1}$  Piret Kuusk,$^{2}$ and Margus Saal$^{3}$

\vspace{0.3cm}
{\it Institute of Physics, University of Tartu,
Riia 142, Tartu 51014, Estonia}

\end{center}

\vspace{1,5cm}

\begin{abstract}
We consider the correspondence between the Jordan frame and the Einstein
frame descriptions of scalar-tensor theory of gravitation. 
We argue that since the redefinition of the scalar field is not differentiable
at the limit of general relativity the correspondence between the two
frames is lost at this limit. 
To clarify the situation we analyse the dynamics of the scalar field
in different frames for two distinct scalar-tensor cosmologies with 
specific coupling functions and
 demonstrate that the corresponding scalar field
phase portraits are not equivalent
for regions containing the general relativity limit.
Therefore the answer to the question 
whether general relativity is an attractor for the theory 
depends on the choice of the frame. 
\end{abstract}

\vspace{1cm}
{\bf PACS}: 98.80.Jk, 04.50.+h, 04.40.Nr

\vspace{6cm}
$^1$ Electronic address: laur@fi.tartu.ee

$^2$ Electronic address: piret@fi.tartu.ee

$^3$ Electronic address: margus@fi.tartu.ee

\newpage

\section{Introduction}

The generalisation of Jordan-Fierz-Brans-Dicke theory of gravitation 
\cite{jordanfierz_bransdicke,weinberg} known as the scalar-tensor theory 
\cite{STT,nordtvedt70,will,faraoni},
where the gravitational interaction is mediated by a scalar field 
together with the usual metric tensor,
appears in various contexts of theoretical physics:
as dilaton gravity in Kaluza-Klein, superstring and supergravity theories,
as the effective description of braneworld models \cite{braneworld}, 
as an equivalent to modified $f(R)$ gravity \cite{fR}, or in attempts
to describe inflation \cite{inflation,gbq} and dark energy \cite{darkenergy}.

The scalar-tensor theory (STT) can be formulated in the Jordan frame, 
where the scalar field $\Psi$ is coupled nonminimally to the Ricci scalar $R$ 
but not directly to the matter, whereas the scalar field
kinetic term involves an arbitrary function $\omega(\Psi)$.
It is possible to 
write the theory in the form reminiscent of the Einstein
general relativity where the scalar field is minimally coupled to the 
Ricci scalar and its kinetic term is in the canonical form. 
In this case the field equations are mathematically less complicated,
but at the price of making the matter couplings dependent on the scalar field.
Going from the Jordan to the Einstein frame proceeds through
two transformations:

1.  A conformal transformation of the Jordan frame metric 
$g_{\mu\nu}$ into the Einstein frame metric $\tilde{g}_{\mu\nu}$;

2. A redefinition of the original scalar field $\Psi$ into $\phi$ 
to give its kinetic term a canonical form.

The problem of physical interpretation and equivalence of these two frames 
has a long history, but discussions have mostly concerned only the role and 
properties of the conformal transformation
(e.g., \cite{dicke, faraoni, faraoni2}).
Much less attention has been paid to the redefinition of the scalar
field used to put its kinetic term in the canonical form. 
The aim of our paper is to caution against the problems stemming 
from this transformation.
The issue is relevant, e.g., in scalar-tensor cosmology 
where one is interested in whether the scalar field naturally evolves to an 
asymptotically constant value, in which case the solutions of STT for $g_{\mu\nu}$ 
can coincide with those of the Einstein general relativity.
In earlier investigations, which were performed in the Jordan frame, 
the main tool was to estimate the late time behaviour of different 
types of solutions \cite{bm, gbq}. Damour and Nordtvedt
\cite{dn} used the Einstein frame to derive a nonlinear
equation for the scalar field decoupled from other variables 
and found that, e.g., in the case of a
flat FLRW model and dust matter there exists an attractor mechanism
taking the solutions of wide class of scalar-tensor theories to the limit of  
general relativity. 
Their approach was generalized to cases of curved FLRW models with 
nonvanishing self-interaction potentials
with the result that in the flat model and dust matter
 the attractor mechanism is not rendered ineffective \cite{skw2}. 
Yet, some authors
\cite{gm, san} have argued under different assumptions,
but still using the Einstein frame, that the attractor mechanism is not generic 
and may be replaced by repulsion.
In the Jordan frame, the main tool of subsequent investigations has been
the construction of viable cosmological models with present state
very near to general relativity, leaving the question of
generality somewhat aside \cite{mim1, mim3, bp}. 

In what follows, our aim is to indicate a possible source of 
these controversies.
The plan of the paper is the following. In section 2 we recall 
a few basic facts
about the scalar-tensor theory
and express some general considerations why 
the scalar field redefinition
is problematic in the general relativity limit.
In section 3 we study two explicit examples, viz.
$2\omega(\Psi)+3=\frac{3}{1-\Psi}$ and $2\omega(\Psi)+3=\frac{3}{|1-\Psi|}$,
and by plotting the phase portraits for the Jordan 
frame $\Psi$ and the Einstein frame $\phi$ demonstrate how
the scalar field dynamics is qualitatively different in different frames.
In section 4 we clarify
why the previous studies of the attractor mechanism in the Einstein frame
have yielded different results. 
We also make some comments on non-minimally coupled STT and 
the weak field limit (PPN).
Finally in section 5 we draw some conclusions, 
in particular, that if the Jordan frame formulation 
is taken to be definitive for 
a scalar-tensor theory 
then the conditions 
for the attractor mechanism towards general relativity
should be reconsidered in the Jordan frame.

\section{General considerations}

Our starting point is the  action  of a general scalar-tensor theory
in the Jordan frame  
  \beq \label{jf4da}
S_{_{\rm J}}  = \frac{1}{2 \kappa^2} \int d^4 x \sqrt{-g}
      	        \left[ \Psi R(g) - \frac{\omega (\Psi ) }{\Psi}
      		\nabla^{\rho}\Psi \nabla_{\rho}\Psi \right] 
                 + S_{m}(g_{\mu\nu}, \chi_m) \,,
\eeq 
where $\nabla_{\mu}$ 
denotes the covariant derivative with respect to the metric 
$g_{\mu\nu}$, 
$\omega(\Psi)$ is a coupling function, 
$\kappa^2$ is the bare gravitational constant and 
$S_{m}$ is the matter part of the action where $\chi_m$ includes all other
fields.
Different choices of the field dependent coupling function $\omega(\Psi)$
give us different scalar-tensor theories.
We assume that $\Psi \in (0, \ \infty)$ or a subset of it and 
$\omega(\Psi) > -\frac{3}{2}$
to keep the effective Newtonian gravitational constant positive 
\cite{nordtvedt70,bp}.
The corresponding field equations for the metric tensor $g_{\mu\nu}$ 
and the scalar field $\Psi$ are given by
\beq \label{jfge}
G_{\mu \nu}(g) = \frac{\kappa^2}{\Psi } T_{\mu \nu}(g) +
\frac{1}{\Psi } \left(  \nabla_{\mu} \nabla_{\nu}\Psi  -g_{\mu \nu} \, 
\Box \Psi \right) 
+\frac{\omega(\Psi )}{\Psi^2} \left( \nabla_{\mu}\Psi \nabla_{\nu}\Psi
-\frac{1}{2} g_{\mu \nu} \, \nabla^{\rho}\Psi \nabla_{\rho}\Psi \right)  \ ,
\eeq
\beq \label{jfPsi}
\Box \Psi = \frac{\kappa^2}{(2\omega(\Psi) + 3)}  \, T(g)
  -\frac{1}{(2\omega(\Psi) +3)} \frac{d\omega}{d\Psi} \nabla^{\mu}\Psi 
\nabla_{\mu}\Psi \,.
\eeq

Although STT and general relativity are mathematically distinct theories,
we may conventionally speak of  ``the general relativity limit of STT" in
the sense of a regime of the solutions of STT where their observational
predictions are identical with those of general relativity.
In typical observational tests of gravitational theories the
parametrized post-Newtonian (PPN) formalism is used for slowly moving 
spherically symmetric systems in the weak field approximation. 
Nordtvedt  \cite{nordtvedt70} has demonstrated that the PPN parameters 
of a STT (with a distinct coupling function $\omega (\Psi)$)  coincide with 
those of general relativity with the Newtonian gravitational constant 
$G_N =\kappa^2/ \Psi_0$ if
\be \label{gen_rel_limit}
\displaystyle\lim_{\Psi\to\Psi_0}\frac{1}{\omega (\Psi)}=0  \,, \qquad  \qquad
\displaystyle\lim_{\Psi\to\Psi_0}\frac{1}{\omega^3(\Psi)}
\frac{d\omega}{d\Psi}=0  \,.
\ee
Let us denote the value
$\Psi =\Psi_0 = const$  as ``the general relativity limit of  STT". 
This definition allows us to call a solution of STT as
``approaching the general relativity limit" if the difference between these 
solutions is asymptotically vanishing.

Upon the conformal rescaling
$\tilde{g}_{\mu\nu} = \Psi \, g_{\mu\nu}$ 
the action (\ref{jf4da}) transforms into 
\beq \label{ef4daPsi} 
S= \frac{1}{2 \kappa^2} \int d^4 x \sqrt{-\tilde{g}} 
\left[ R(\tilde{g}) - \frac{ (2 \omega + 3)}{2 \Psi^2} \, 
\tilde{g}^{\mu\nu} \, \tilde{\nabla}_{\mu} \Psi \tilde{\nabla}_{\nu} \Psi \right]
+ S_{m} ( \Psi^{-1} \tilde{g}_{\mu\nu}, \chi_m )\,, 
\eeq 
where $\tilde{\nabla}_{\mu}$ denotes the covariant derivative 
with respect to the metric $\tilde{g}_{\mu\nu}$. 
The kinetic term of the scalar field obtains the canonical form
by the means of a field redefinition
\be \label{ruut}
2 (d\phi)^2 = \frac{(2\omega + 3)}{2 \Psi^2} (d\Psi)^2 \,,
\ee
that determines a double-valued correspondence 
\be \label{redef}
\frac{d\Psi}{d\phi} = \mp \frac{2 \Psi}{\sqrt{2\omega(\Psi) + 3}}\,.
\ee
This double-valuedness may be interpreted as defining two distinct Einstein 
frame theories which correspond to a Jordan frame theory,
i.e., we may choose one of the two possible signs and keep it throughout all
subsequent calculations.
But in the literature one also meets another approach, where the scalar field is 
allowed to evolve from one branch (sign) to another.
In order to fully clarify the issue we retain the possibility of both signs.

The resulting Einstein frame action is given by
\beq \label{ef4da}
S_{_{\rm E}}= \frac{1}{2 \kappa^2} \int d^4 x \sqrt{-\tilde{g}}
\left[ R(\tilde{g}) - 2 \tilde{g}^{\mu\nu} \, \tilde{\nabla}_{\mu} \phi 
\tilde{\nabla}_{\nu} \phi \right] 
+ S_{m}( \Psi^{-1}(\phi) \tilde{g}_{\mu\nu}, \chi_m )\,,
\eeq
where the range of $\phi$ depends on the range of coupling function $\omega(\Psi)$
as given by Eq. (\ref{redef})  and can be determined only upon choosing a 
particular function $\omega(\Psi)$.
The corresponding field equations are
\beq
G_{\mu\nu}(\tilde{g}) = \kappa^2  T_{\mu\nu}(\tilde{g})  +
2 (\tilde{\nabla}_{\mu} \phi \tilde{\nabla}_{\nu} \phi - \frac{1}{2}
g_{\mu\nu} \tilde{\nabla}^{\rho} \phi  \tilde{\nabla}_{\rho}\phi ) \,,
\eeq
\beq
\tilde{\Box} \phi =
\frac{\kappa^2}{2} \alpha(\phi) \, T(\tilde{g}) \,,
\eeq
where
\beq
T_{\mu \nu}(\tilde{g}) = - \frac{2}{\sqrt{-\tilde{g}}}
\frac{\delta S_{m}( \Psi^{-1}(\phi) \tilde{g}_{\mu\nu}, \chi_m ) }{\delta 
\tilde{g}^{\mu \nu}} \,, \quad
\label{ef_conservation}
\tilde{\nabla}^{\mu}  T_{\mu \nu}(\tilde{g})  =  - \alpha(\phi)  T(\tilde{g})   \tilde{\nabla}_{\nu} \phi \ ,
\eeq 
and
\beq 
\label{alphadef}
\alpha(\phi) = \sqrt{\Psi} \frac{d (\sqrt{\Psi})^{-1}}{d\phi} = \pm \frac{1}{\sqrt{2 \omega(\Psi(\phi)) +3}} \, .
\eeq
``The limit of general relativity" corresponding to Eq. (\ref{gen_rel_limit}) 
is now given by $\phi = \phi_0$, satisfying  $\alpha(\phi_0) =0$.

The mathematical form of the scalar field redefinition (\ref{redef})
and of the ensuing Eq. (\ref{alphadef}) raise two concerns here.

1. The property of double-valuedness of $\phi(\Psi)$ is generally harmless,
simply meaning that the original Jordan frame physics is represented by two 
equivalent copies in the Einstein frame description 
(related by $\phi \leftrightarrow -\phi$).
However, these two copies meet at the point $\Psi_0$ 
corresponding to the limit of
general relativity (\ref{gen_rel_limit}). 
Since $d\Psi/d\phi$ vanishes there, 
this point has to be a point of inflection or a local extremum 
of function $\Psi (\phi)$ (for an illustration see Fig. 1). 
The former case corresponds to picking the same sign
in Eq. (\ref{redef}) on both sides $\Psi<\Psi_0$ and $\Psi>\Psi_0$,
while in the latter case the derivative 
$d\Psi/d\phi$ changes sign, which occurs with changing 
the sign in Eq.  (\ref{redef}). 
The second option remains the only possibility when the scalar 
field in the Jordan frame is assumed to have a restricted domain and 
$\Psi_{0}$ resides at its boundary.
It turns out that the choice of the domain of $\Psi$ and related issue of signs in Eq. (\ref{redef})
are significant and in section 4 we discuss  how different assumptions
yield qualitatively different results in the Einstein frame, namely, 
whether $\phi_0$ is a generic attractor or not.

2. The property of $d\Psi/d\phi$ to vanish at $\Psi_0$ implies
that as the field $\Psi$ reaches the value $\Psi_0$ 
its dynamics as determined by the variational principle loses the 
correspondence with the dynamics of $\phi$. 
Indeed, an infinitesimal variation of an action
functional is invariant at a regular change of variables, 
so the variation of STT action functional can be given
in two different forms
\be
\delta S = \frac{\delta S_J}{\delta \Psi} \delta \Psi +
  \frac{\delta S_J}{\delta g_{\mu \nu}} \delta g^{\mu \nu} =
  \frac{\delta S_E}{\delta \phi} \delta \phi +
  \frac{\delta S_E}{\delta {\tilde g}_{\mu\nu}} \delta {\tilde g}^{\mu\nu} \,.
\ee
But this relation may not hold if estimated at 
extremals ($\Psi_{0}$, $g_{\mu\nu}$), 
since $\delta \phi = \frac{d\phi}{d\Psi} \delta \Psi$ and
$\frac{d\phi}{d\Psi}$ diverges
there according to Eq. (\ref{redef}), 
i.e., the change of variables is not regular.

Here  a remote analogy with coordinate patches in topologically nontrivial spaces 
suggests itself. 
For instance, if we describe   particle's worldlines in terms of Schwarzschild coordinates
we can not go beyond the $r = 2m$ ``boundary", however, if we use
Kruskal coordinates we would be able to follow the particle's world line beyond it. 
In the case of scalar-tensor theories, the choice of ``field coordinates" could also entail
similar effects. 
Yet, invariant description of STT in field space is still not well understood (e.g., \cite{sfl}).

\section{Examples}

\subsection{$2\omega(\Psi) +3 = \frac{3}{(1 - \Psi)} $}

Let us consider a scalar-tensor cosmology 
with 
the  coupling function
\be \label{j_cf}
\omega(\Psi) = \frac{3}{2} \frac{\Psi}{(1 - \Psi)}   \,,
\ee
with a restricted domain $\Psi \in (0,\ 1]$,
which arises as an effective description of Randall-Sundrum two-brane
cosmology \cite{ks2,meie2}, and has also been
considered before as an example of 
conformal coupling \cite{conformal,gbq,mim3}
or as a STT with vanishing scalar curvature \cite{gerard}.
The field equations for a flat Universe ($k=0$)
with the FLRW line element and perfect barotropic fluid matter,
$p = (\Gamma - 1) \rho $, read 
\beq 
\label{00}
  H^2 &=& 
- H \frac{\dot \Psi}{\Psi} 
+ \frac{1}{4} \frac{\dot \Psi^2}{\Psi (1 - \Psi)} 
+ \frac{\kappa^2}{3} \frac{ \rho}{\Psi} \,, 
\\ \nonumber \\
\label{mn}
2 \dot{ H} + 3 H^2 &=& 
- 2 H \frac{\dot{\Psi}}{\Psi} - \frac{3}{4} 
\frac{\dot{\Psi}^2}{\Psi(1-\Psi)} 
- \frac{\ddot{\Psi}}{\Psi} 
-\frac{\kappa^2}{\Psi} (\Gamma-1)\rho \,, 
\\ \nonumber \\
\label{deq}
\ddot \Psi &=& - 3H \dot \Psi - \frac{1}{2} \frac{\dot \Psi^2}{(1-\Psi)} 
+ \frac{\kappa^2}{3}(1-\Psi) \ (4-3 \Gamma)  \rho  \,
\eeq 
($H \equiv \dot{a} / a$), while the conservation law is the usual
\beq \label{j_cl_A}
\dot{\rho} + 3 H \Gamma \rho = 0 \,.
\eeq
The limit of general relativity  (\ref{gen_rel_limit}) is reached at $\Psi \rightarrow 1$.
Eqs. (\ref{00})--(\ref{deq}) are singular at this value, however, as we see soon, 
it corresponds to a fixed point in a dynamical system describing the scalar field.

The Einstein frame description is obtained by 
conformally rescaling the metric,
$\tilde{g}_{\mu\nu} = \Psi \, g_{\mu\nu}$,
followed by a coordinate transformation to keep the FLRW form of
the line element,
\beq
\tilde{a} = \sqrt{\Psi} a \,, \qquad d\tilde{t} = \sqrt{\Psi} dt \,,
\qquad \tilde{\rho} = \Psi^{-2} \rho \,.
\eeq
The redefinition (\ref{redef}) of
the scalar field which gives its kinetic term the usual canonical form,
\beq \label{re_def}
\frac{d \phi}{d \Psi} =\mp \sqrt{\frac{3}{4} \, \frac{1}{\Psi^2\, 
( 1- \Psi)}} \,,
\eeq
\begin{figure}
\begin{center}
\hspace{-20mm}
\includegraphics[angle=0,width=80mm]{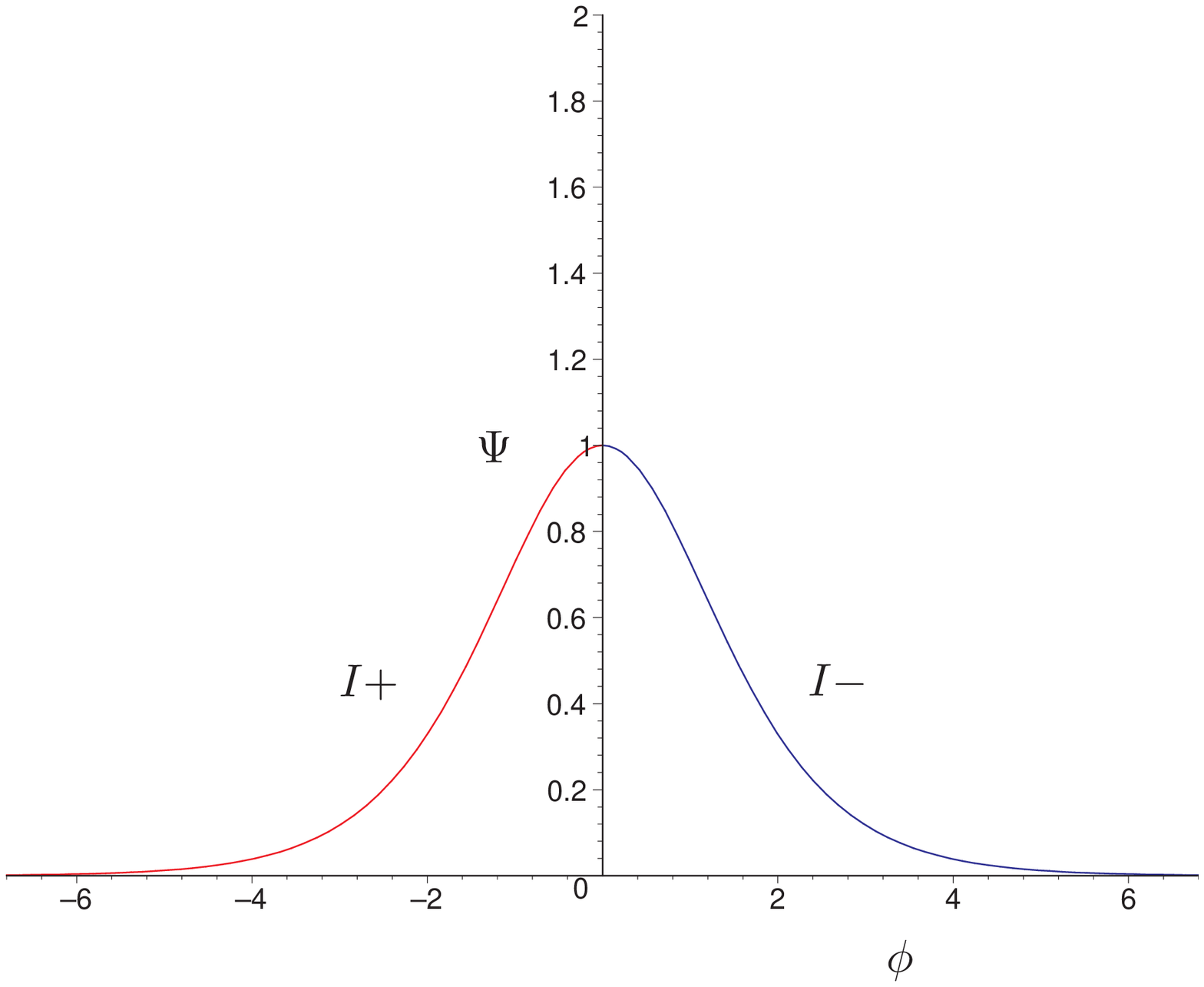}
\includegraphics[angle=0,width=80mm]{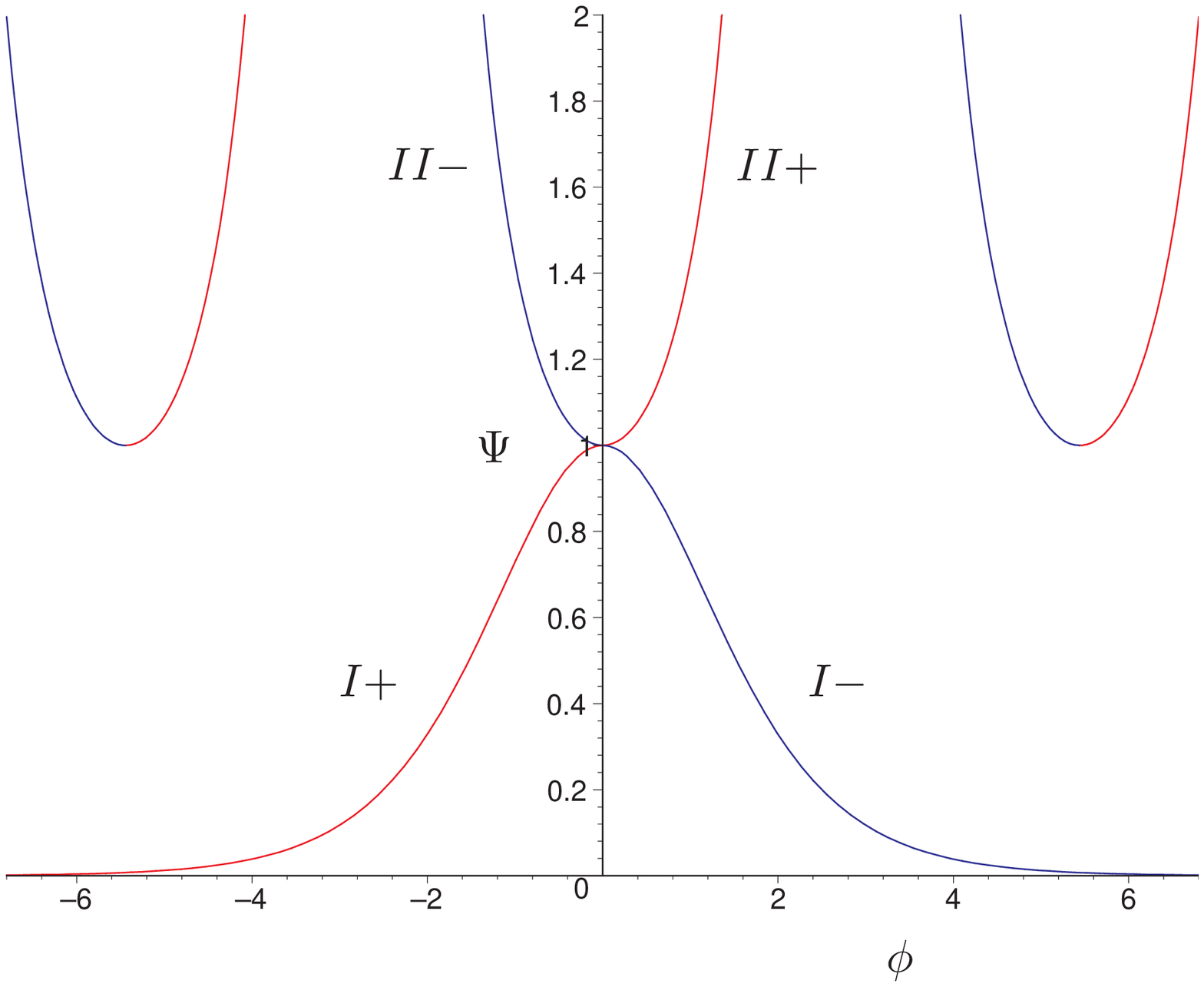}
\caption{{\sl{Solution of the scalar field redefinition (\ref{redef}) in the 
example 1) $2\omega(\Psi)+3=\frac{3}{1-\Psi}$ (left), and 2) $2\omega(\Psi)+3=\frac{3}{|1-\Psi|}$ (right).}}}
\end{center}
\end{figure}
is solved by 
\beq \label{ef_psitophi}
\pm \phi  =    \sqrt{3} \ {\rm arctanh} (\sqrt{1-\Psi}) \, , 
\qquad \qquad 
\pm \sqrt{1 - \Psi} =  {\tanh}\left( \frac{\phi}{\sqrt{3}} \right)\,.
\eeq
The solution is plotted on Fig. 1 left. 
There are two branches $I+$ and $I-$ corresponding to the positive and negative
signs in Eq. (\ref{re_def}) respectively. The map $\Psi \rightarrow \phi$ is 
double valued, the two branches $\phi \in (-\infty, 0]$ and  
$\phi \in (\infty, 0]$ 
define two Einstein frame copies of the Jordan frame physics of 
$\Psi \in (0,1]$. The two branches meet at the point $\phi_0 = 0$,
which corresponds to the limit of general relativity, $\Psi_0 =1$.
For this point there is a choice to be made with two options: 
either we allow $\phi$ to pass from one branch to another, or not.
The first option would mean that $\phi$ can jump 
from one copy of the Einstein frame
description to another equivalent copy. 
In the Jordan frame description this corresponds
to $\Psi$ bouncing back from $\Psi_0$. 
The second option would mean that the evolution of $\phi$ 
has to end at $\phi_0$ even when it reaches this point
with non-vanishing speed.
Of course, there would be no problem, if the equations for $\phi$ 
were already ``aware'' of this and 
never allowed $\phi$ to reach $\phi_0$ with non-vanishing speed.
Unfortunately this is not so, as we will see in the following.

The Einstein frame equations read
\beq  \label{ef_00}
\tilde{H}^2 
&=& \frac{1}{3}\, \dot{\phi}^2 + 
\frac{\kappa^2}{3} \,\tilde{\rho} \, \\
\label{ef_mn}
2 \dot{\tilde{H}} +3 \tilde{H}^2
&=& - \dot{\phi}^2 - \kappa^2 (\Gamma-1)\tilde{\rho} \ , \\
\label{ef_deq}
\ddot{\phi} + 3 \tilde{H} \, \dot{\phi} 
&=& - \frac{1}{2} \kappa^2 \, \alpha (\phi) \,(4-3\Gamma) \tilde{\rho} \,, \\
\label{ef_cons}
\dot{\tilde{\rho}} + 3 \tilde{H} \Gamma \ \tilde{\rho} 
&=& \alpha(\phi) \, (4-3\Gamma)\tilde{\rho}
 \ \dot{\phi}\,. 
\eeq
Here
\be \label{alpha}
\alpha(\phi)
  =  \frac{1}{\sqrt{3}}\ {\rm tanh}\ \left(\frac{\phi}{\sqrt{3}} \right) 
\ee
acts as a coupling function in the wave equation 
for the scalar field (\ref{ef_deq}) and also occurs in the matter 
conservation law (\ref{ef_cons}). The limit of general relativity,
$\alpha(\phi_0)=0$,
is at $\phi_0 = 0$.

In the following let us consider the case of dust matter ($\Gamma = 1$).
 Eqs.~(\ref{00})--(\ref{j_cl_A}) and (\ref{ef_00})--(\ref{ef_cons})
can be numerically integrated (Fig. 2). The result explicitly 
supports the concern that the dynamics of the scalar field can be different
in different frames when the limit of general relativity is reached:
while the Jordan frame solution
converges to the limit of general relativity ($\Psi_0=1$), 
the Einstein frame solution of the same initial conditions 
(properly transformed from the Jordan frame)
evolves through the corresponding point ($\phi_0=0$). 
Here we allowed $\phi$ to jump from the branch $I-$ to the branch $I+$,
since otherwise it must have been stopped abruptly at $\phi_0=0$, which is
not in
accordance with Eqs. (\ref{ef_00})--(\ref{ef_deq}). 
To confirm that 
this difference in the behaviour of the Jordan and the Einstein frame
descriptions is not due to numerical effects,
but is truly encoded in the dynamics, we have to
look at the phase portraits \footnote{The phase space dynamics
of scalar-tensor cosmology has been studied in some special cases 
\cite{phasespace} and 
used to reconstruct the STT coupling and potential 
by demanding a background $\Lambda$CDM cosmology \cite{cnp}, while
general considerations about the phase space geometry
were given by Faraoni \cite{faraoni3}.
However, our approach here is focused upon the phase space of
the decoupled equation for the scalar field \cite{meie2}.}.
\begin{figure}
\begin{center}
\hspace{-20mm}
\includegraphics[angle=0,width=87mm]{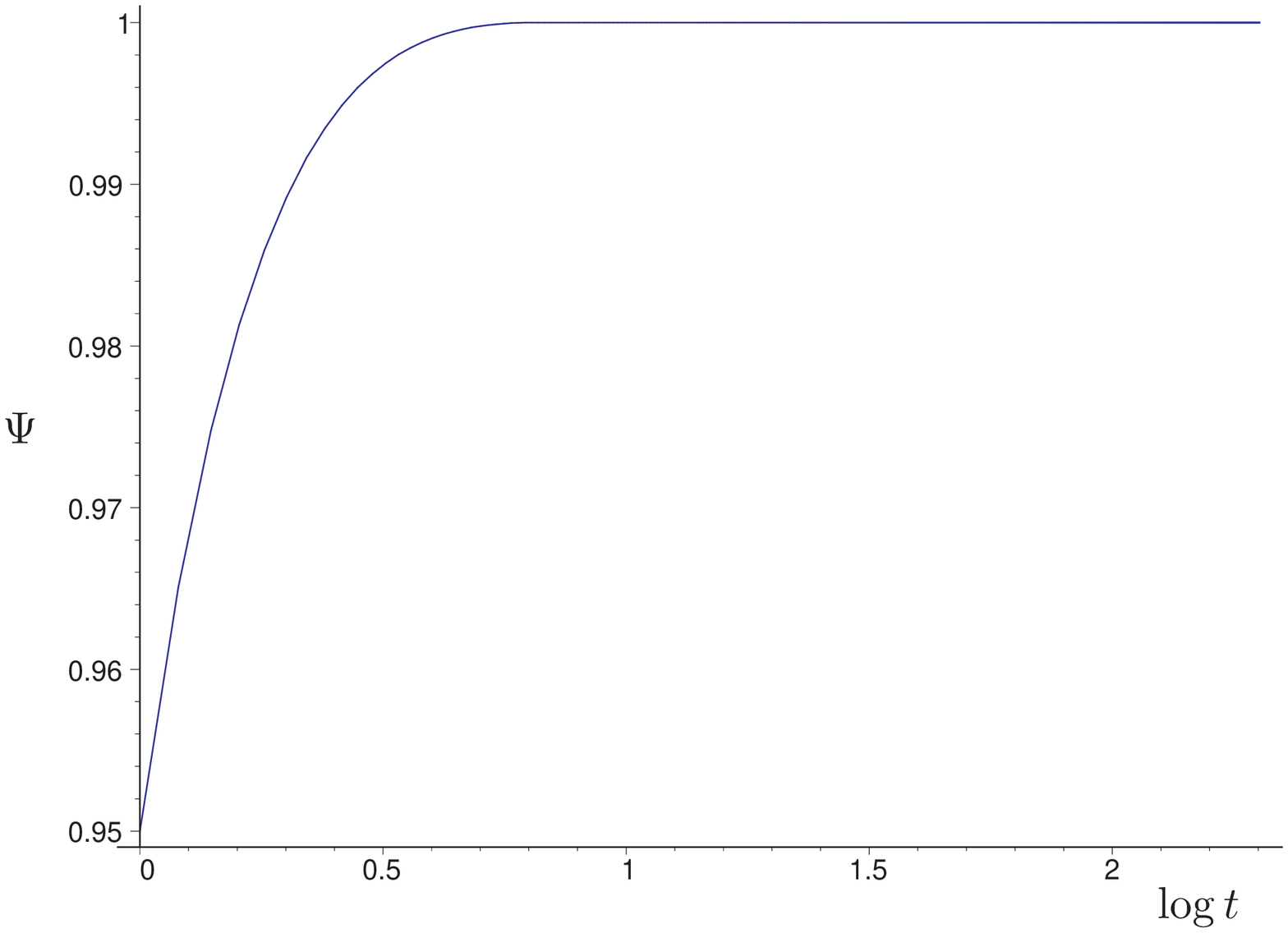}
\includegraphics[angle=0,width=87mm]{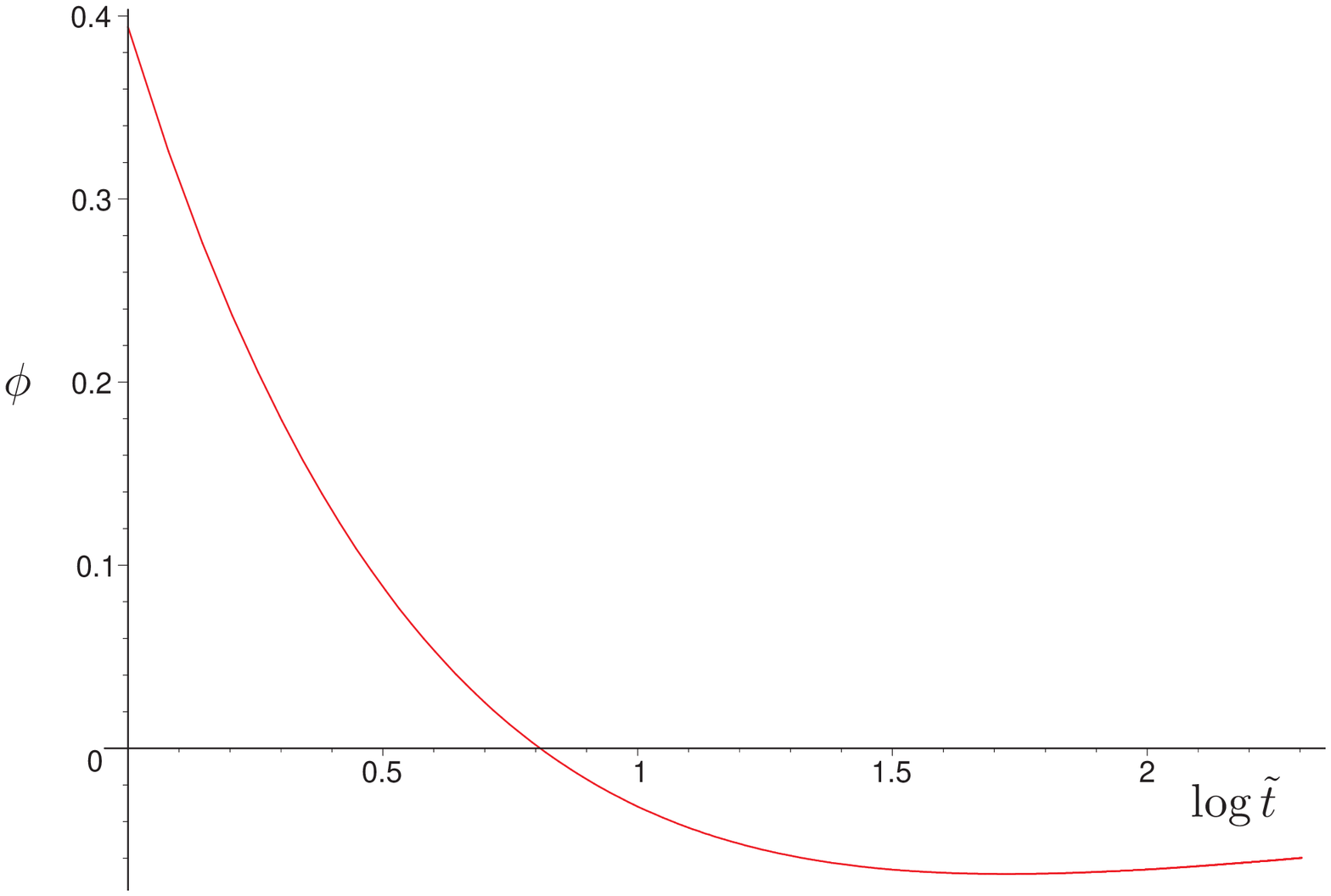}
\caption{{\sl Numerical solution of example 1) 
with the initial condition $\Psi(0)=0.95$, 
$\dot{\Psi}(0)=0.095$, $\rho(0)=1$, $a(0)=1$ in the Jordan frame (left) and 
Einstein frame (right). Note that since $\Psi \approx 1$ 
the respective time variables 
$t$ and $\tilde t$ differ only slightly.}}
\end{center}
\end{figure}

By a change of variables introduced by Damour and Nordtvedt \cite{dn} 
it is possible to combine the field equations to get a dynamical 
equation for the scalar field which does not manifestly 
contain the scale factor or matter density. 
In the Jordan frame this amounts to defining a new time variable
\cite{san}
\be \label{jt_ptime}
dp = h_c dt \equiv \left( H + 
\frac{\dot{\Psi}}{2\Psi} \right) dt \,.
\ee
Then from Eqs. (\ref{00})--(\ref{deq})
the following ``master'' equation for the scalar field can be 
derived \cite{san, meie2}:
\be \label{mejf}
8 (1-\Psi)\frac{\Psi''}{\Psi} - 3 \left( \frac{\Psi'}{\Psi}
\right)^3
-2 ( 3-5\Psi) \left(\frac{\Psi'}{\Psi}\right)^2 
+ 12 (1-\Psi)\frac{\Psi'}{\Psi} - 8(1-\Psi)^2 
= 0 \,,
\ee
where primes denote the derivatives with respect to $p$.
The Friedmann constraint (\ref{00}) in terms of the new time variable $p$ 
can be written as
\be \label{jf_ptime_friedmann}
h_c^2 = \frac{\kappa^2 \ \rho}
{3\Psi \left( 1- \frac{{\Psi'}^2}{4 \Psi^2(1-\Psi)} \right)} \; .
\ee
Assuming that $\rho$ is positive definite, 
the constraint restricts the dynamics to explore only the region
\be \label{rho Friedmann allow}
|\Psi'| \leq | 2 \Psi \sqrt{1-\Psi}| \; .
\ee
Notice, Eq. (\ref{jf_ptime_friedmann}) assures that the time 
reparametrisation (\ref{jt_ptime}) works fine, as within the borders of
the allowed phase space $p$-time and $t$-time always run in the same
direction. Also, from $\dot{\Psi} = h_c \Psi'$ it is easy to see that 
$\dot\Psi = 0$ corresponds to $\Psi' = 0$, while $\dot{\Psi} = \pm \infty$
corresponds to the boundary $\Psi' = \pm 2 \Psi \sqrt{1-\Psi}$.

Let us introduce variables $x \equiv \Psi$, $y \equiv \Psi'$ and write 
Eq. (\ref{mejf}) as a dynamical system
\beq \label{mejf_dynsys}
\left \{ \begin{array}{rcl}
x' & = & y  \\
y' & = & \frac{3 y^3}{8 x^2 (1-x)} + \frac{(3-5x)y^2}{4 x (1-x)}
-\frac{3 y}{2} + x (1-x) \, .
\end{array}
\right.
\eeq
\begin{figure}
\begin{center}
\hspace{-5mm}
\includegraphics[angle=-90,width=75mm]{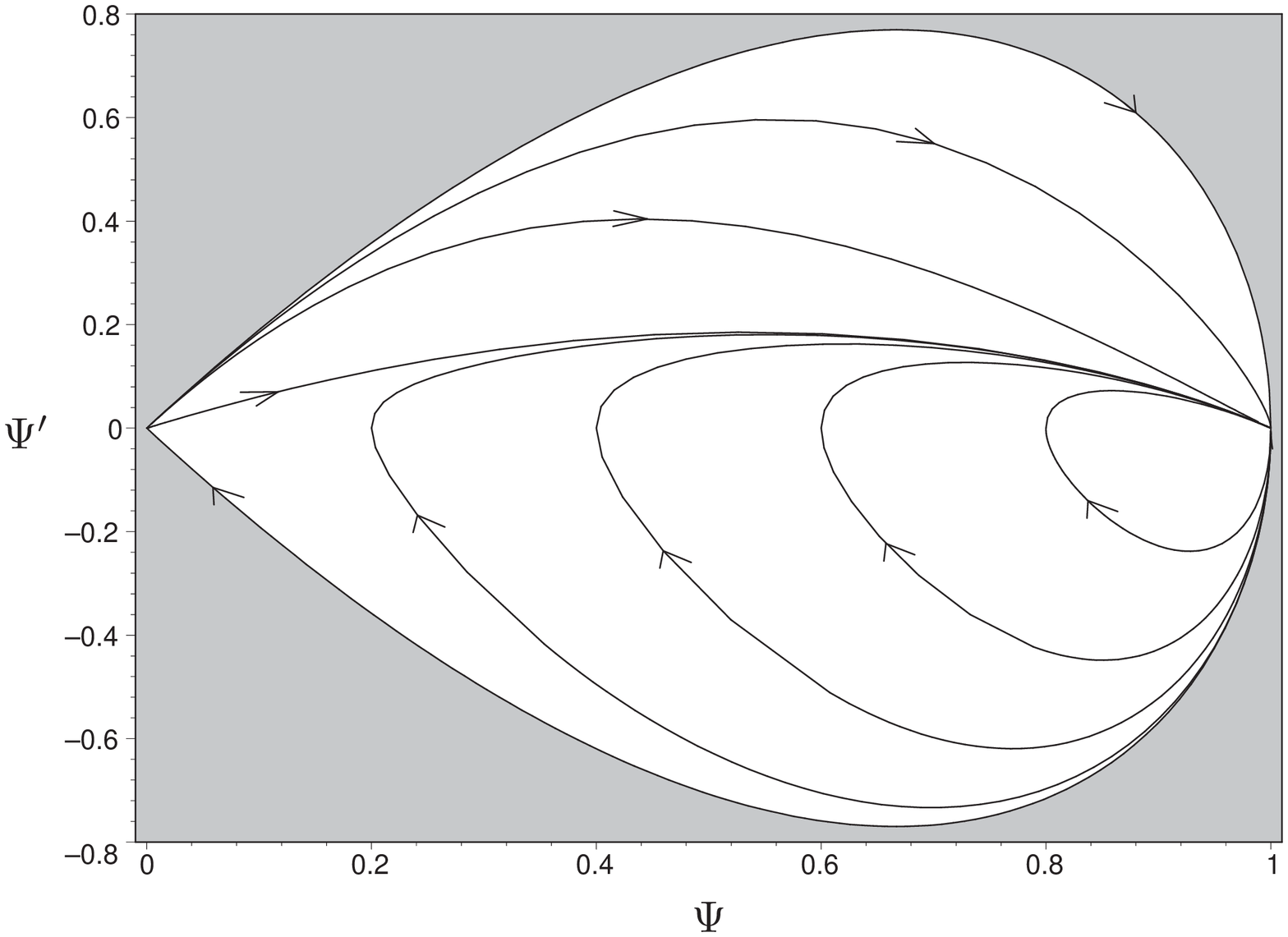}
\hspace{1cm}
\includegraphics[angle=-90,width=75mm]{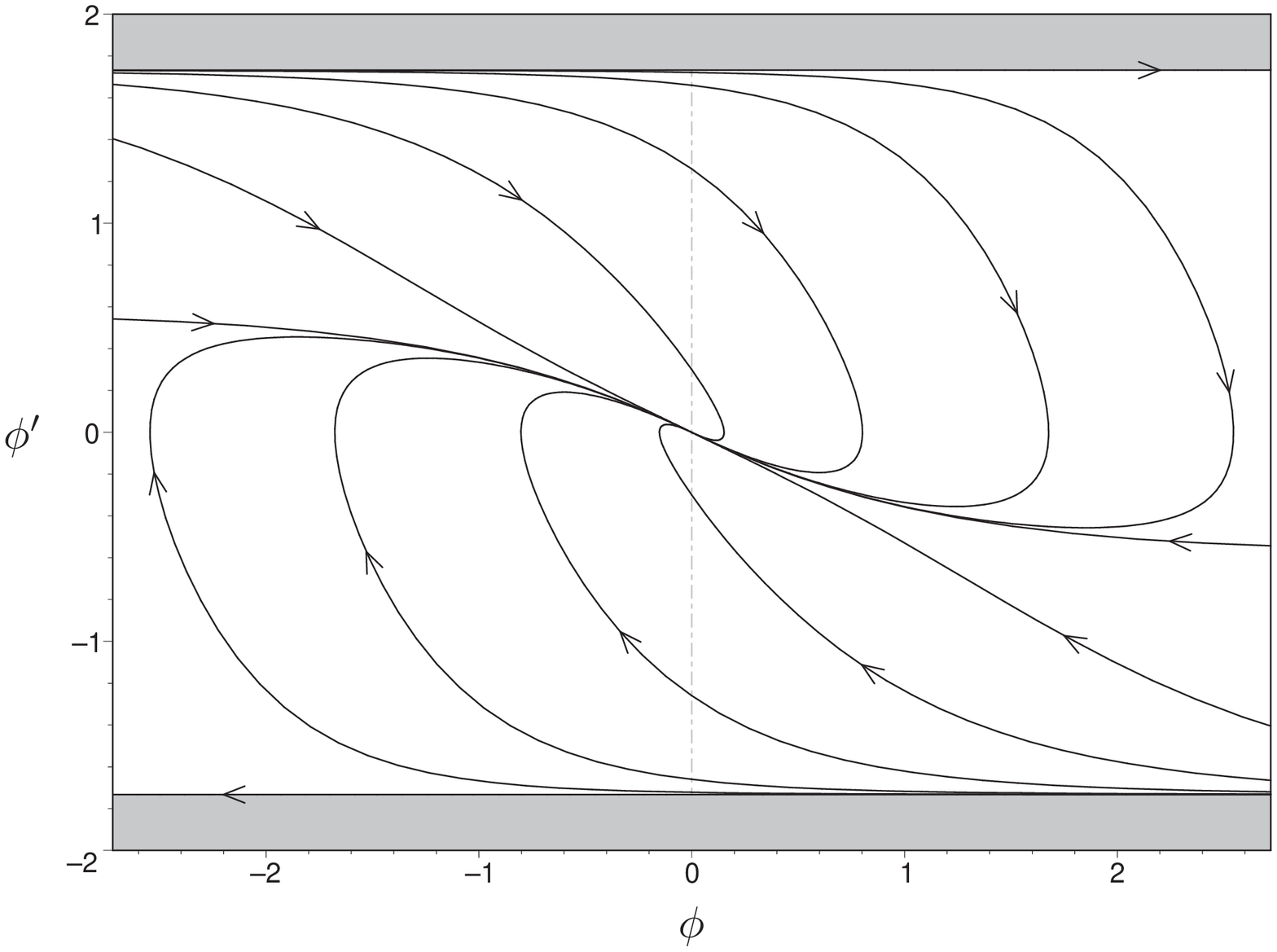}
\caption{{\sl{Example 1) 
phase portraits of the scalar field ``master'' equation 
(\ref{mejf}) in the Jordan frame (left) and its analogue (\ref{medn}),
(\ref{alpha}) 
in the Einstein frame (right).}}}
\end{center}
\end{figure}
There are two fixed points:
\bi
\item a saddle point at $(x=0, y=0)$, with repulsive and attractive
eigenvectors tangential
to the upper and lower boundaries  $y = \pm 2x \sqrt{1-x}$, respectively,
\item a spiralling attractor at $(x=1, y=0)$, but notice here
the trajectories also need to respect 
the boundaries of the allowed region.
\ei
As can be see from the phase portrait (Fig. 3 left) 
all trajectories begin in the infinitesimal 
vicinity of one of the two fixed points. 
Also all trajectories are collected by the attractor, except for the
marginal trajectory along the boundary $y = -2x \sqrt{1-x}$,
which runs into the saddle point. Translating back to the original time
$t$ it turns out that the attractor
corresponds to the limit of general relativity ($\Psi \rightarrow 1,
\dot\Psi \rightarrow 0$) for all trajectories within the allowed phase space.

In the Einstein frame the new time variable is given by \cite{dn,san}
\be
dp = \tilde{H} \, d \tilde{t} \, ,
\ee
and from Eqs. (\ref{ef_00})--(\ref{ef_deq}) follows
an analogous ``master'' equation 
\be \label{medn}
\frac{2}{3-\phi'^2}\, \phi'' +  \, \phi' 
= - \alpha (\phi) \,,
\ee
where primes denote the derivatives with respect to $p$
and $\alpha(\phi) $ is given by Eq. (\ref{alpha}). Now
the allowed phase space is constrained by 
\be \label{ef_dynsys_allow}
\phi' \leq \pm \sqrt{3} \,,
\ee
$\dot\phi=0$ corresponds to $\phi'=0$, while $\dot\phi= \pm \infty$
corresponds to the boundary $\phi'= \pm \sqrt{3}$.
In the variables $x \equiv \phi$, $y \equiv \phi'$ 
Eq. (\ref{medn}) reads 
\beq \label{meef_dynsys}
\left \{ \begin{array}{rcl}
x' & = & y  \\
y' & = & - y \ (3 - y^2) - \frac{(3 - y^2)}{\sqrt{3}}\ {\rm tanh} 
\left( \frac{x}{\sqrt{3}} \right) \,.
\end{array}
\right. 
\eeq
There is one fixed point:
\bi
\item an attractor at $(x=0, y=0)$.
\ei
As can be observed from the phase portrait (Fig. 3 right) the attractor 
collects all the trajectories, except the marginal ones which run along the 
boundaries.

Despite both cases exhibiting an attractor behaviour, the Jordan and Einstein
frame phase portraits are not equivalent.
The Einstein frame portrait is symmetric under
$x \leftrightarrow -x, y \leftrightarrow -y$, related to
the two branches (two copies) discussed above. 
The transition form one branch to another is smooth and there is no constraint
on the Einstein frame dynamics to prevent the trajectories from passing
through $\phi=0$. 
In fact, all the Einstein frame trajectories
do cross once from one branch to another, except for the two trajectories which
flow directly from $\phi = \pm \infty$ to the fixed point.
This general behaviour confirms that the Einstein frame 
solution on Fig. 2 right 
does indeed evolve through $\phi=0$ and the crossing 
is not an artifact of numerical errors in a sensitive region.
However, the passing of $\phi$ from one branch to another 
would in the Jordan frame description correspond to $\Psi$
evolving to $\Psi=1$ and then bouncing back to $\Psi<1$.
This does not happen, as is illustrated by the solution on
Fig. 2 left, which monotonously converges to $\Psi=1$. 
The analysis of the Jordan frame phase portrait makes it completely clear.
No trajectory does change from $\Psi'>0$ to $\Psi'<0$, 
all trajectories with $\Psi'>0$ necessarily flow towards $\Psi=1$,
and $\Psi=1$ is a fixed point, i.e, there is no way back.

An alternative option would be to cut the Einstein frame 
phase portrait along $\phi=0$ into two copies and maintain both 
separately. Then there would be no problematic 
crossing from one branch to another,
however, in this case there is a mismatch between 
the extent of the past or future of the solutions in different
frames.
All generic Einstein frame solutions either terminate at finite
time (run to $\phi=0$ with $\phi' \neq 0$) or begin at finite
time (emerge at $\phi=0$ with $\phi' \neq 0$). 
Yet, all Jordan frame solutions have infinite past and 
infinite future (they begin near a fixed point and run into a fixed point).
On Fig. 2 this would correspond to terminating the Einstein frame
solution at $\phi=0$ at a finite time, while its Jordan frame counterpart
can enjoy an infinite time in approaching $\Psi=1$.

The reason for the incompatibility of the Jordan and Einstein frame 
pictures lies, of course, in the singular behaviour 
of the transformation (\ref{re_def}) at $\phi =0$, which 
maps the point $(\Psi = 1, \Psi' = 0)$ in the Jordan frame to the 
whole line $(\phi = 0, |\phi'|  < \sqrt{3})$ on the Einstein
frame phase diagram. 
The Jordan frame solutions which approach $\Psi \rightarrow 1$
with $\Psi' \rightarrow 0$ get mapped to the Einstein frame solutions 
$\phi \rightarrow 0$ with
arbitrary $\phi'$ which therefore do not necessarily stop at $\phi=0$, 
but can evolve though.
This is a manifestation of our general observation
that at the limit of general relativity 
the dynamics of the Einstein frame $\phi$ loses any correspondence with the 
dynamics of the Jordan frame $\Psi$. 
The fact that the Einstein frame description involves two copies of the 
Jordan frame physics and the problem whether or not to glue these 
copies together really becomes an issue since the $\phi$ trajectories 
lose correspondence with the $\Psi$ trajectories at this point.
None of the two options on how to deal with the two branches yields an
acceptable result.


\subsection{$2 \omega(\Psi)+ 3 = \frac{3}{|1 - \Psi|} $}

As a second example, let us consider a scalar-tensor cosmology 
with the coupling function 
\be \label{jm_cf}
\omega(\Psi) = \frac{3}{2} \ \frac{1 - |1 -\Psi|}{|1 - \Psi|}  \,,
\qquad  \Psi \in (0, +\infty) \,,
\ee
which
belongs to subclasses (a) and (c) in the classification
proposed by Barrow and Parsons \cite{bp} and was studied
before by Serna et al \cite{san}.
The field equations for a flat Universe ($k=0$)
with the FLRW line element and perfect fluid matter now read 
\beq 
\label{00_abs}
H^2 &=& 
-  H \frac{\dot \Psi}{\Psi} 
+ \frac{1}{4}  \ \frac{1 - |1 - \Psi|}{|1-\Psi|}
  \Bigl(\frac{\dot \Psi}{\Psi}\Bigr)^2
+ \frac{\kappa^2}{3} \ \frac{\rho}{\Psi} \,, 
\\ \nonumber \\
\label{mn_abs}
2 \dot{H} + 3 H^2 &=& 
 - 2 H \frac{\dot{\Psi}}{\Psi} - \frac{3}{4} \
  \frac{1 -|1-\Psi|}{|1-\Psi|} \Big(\frac{\dot{\Psi}}{\Psi}\Bigr)^2
- \frac{\ddot{\Psi}}{\Psi}
-\frac{\kappa^2}{\Psi} (\Gamma-1)\rho 
 \,, 
\\ \nonumber \\
\label{deq_abs}
\ddot \Psi &=& - 3H \dot \Psi - 
\frac{1}{2} \ \frac{\dot \Psi^2}{(1-\Psi)} 
+ \frac{\kappa^2}{3}|1-\Psi| \ (4-3\Gamma)\rho  \,.
\eeq
In the case of dust ($\Gamma =1$) 
an analogue of the ``master'' equation (\ref{mejf}) is given by 
\beq \label{mejfm_abs}
8 \ |1-\Psi| \ \frac{\Psi''}{\Psi} - 3 \left( \frac{\Psi'}{\Psi}\right)^3
-2 \ \frac{|1 - \Psi|}{(1 - \Psi)} \ (3-5\Psi)
  \left(\frac{\Psi'}{\Psi}\right)^2
+ 12 \ |1-\Psi| \ \frac{\Psi'}{\Psi} - 8(1-\Psi)^2 
= 0 ,
\eeq
while the Friedmann equation constrains the dynamics to explore
the region
\be \label{Friedmann_allow_abs}
|\Psi'| \leq | \; 2 \Psi \sqrt{|1-\Psi|} \;| 
\ee
only. We can write Eq. (\ref{mejfm_abs}) as a dynamical system
and study the respective phase portrait as before,
see Fig. 4 left. The phase portrait in the region $\Psi \leq 1$
is identical with the previous case (Fig. 3 left), while the region 
$\Psi \geq 1$ is now a new feature. These two regions meet at the point 
$(\Psi=1, \Psi'=0)$, which is also a fixed point. 
It turns out that this fixed point has
different properties for the regions $\Psi \leq 1$ and 
$\Psi \geq 1$.
For the trajectories in the region $\Psi \leq 1$ it functions as a
spiralling attractor as we learned before. For the trajectories in the 
$\Psi \geq 1$ region, however, it is a saddle point with attractive
and repulsive eigenvectors tangential to the lower and upper boundaries 
(\ref{Friedmann_allow_abs}), respectively. Therefore all generic trajectories 
in the $\Psi \geq 1$ region start at $\Psi=\infty$, come arbitrarily 
close to $\Psi=1$ but get turned around and run back to $\Psi=\infty$.
It is not possible for the trajectories to pass from the region 
$\Psi \leq 1$ to the region $\Psi \geq 1$, or vice versa.
\begin{figure}
\begin{center}
\hspace{-5mm}
\includegraphics[angle=-90,width=75mm]{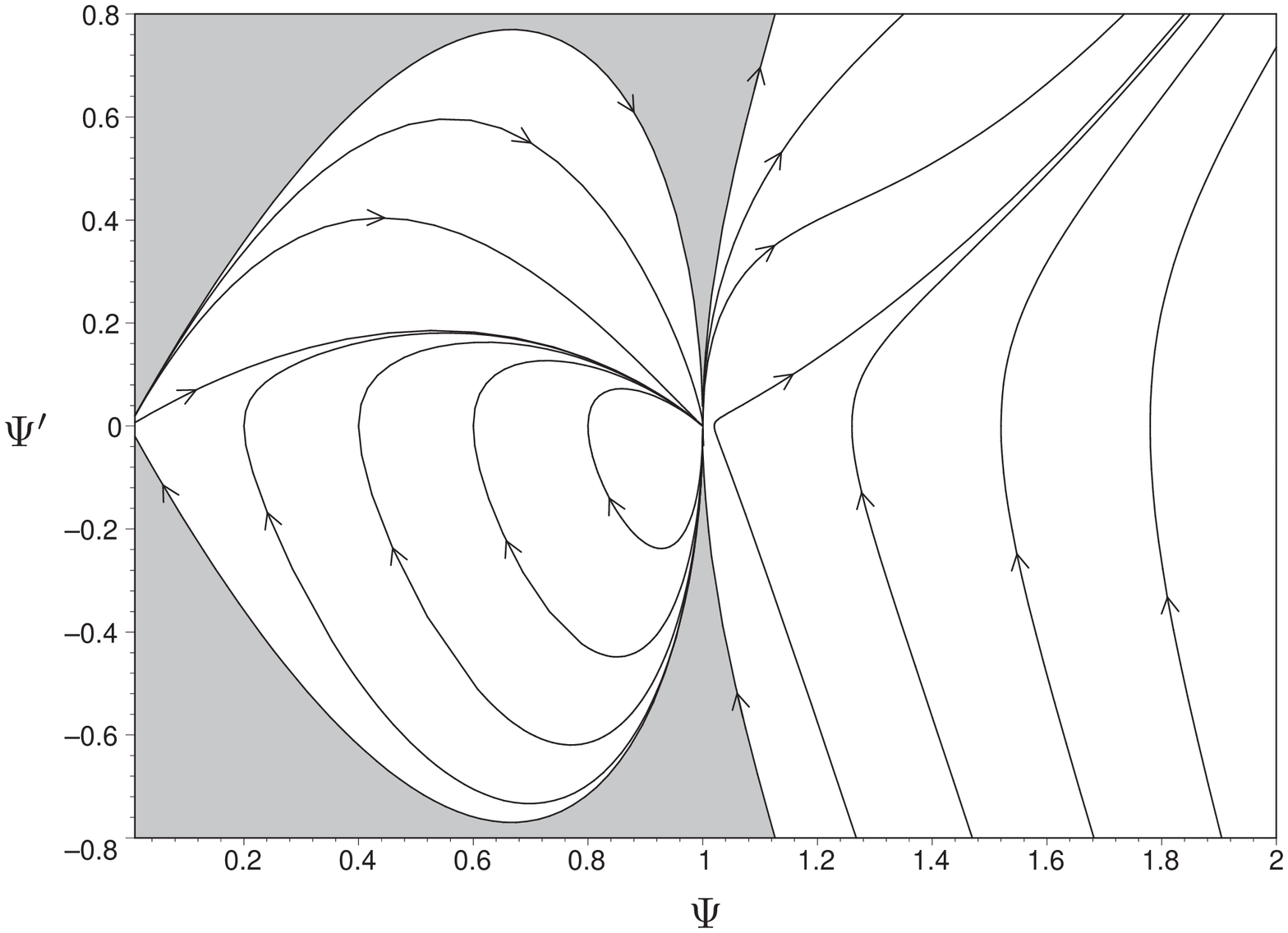}
\hspace{1cm}
\includegraphics[angle=-90,width=75mm]{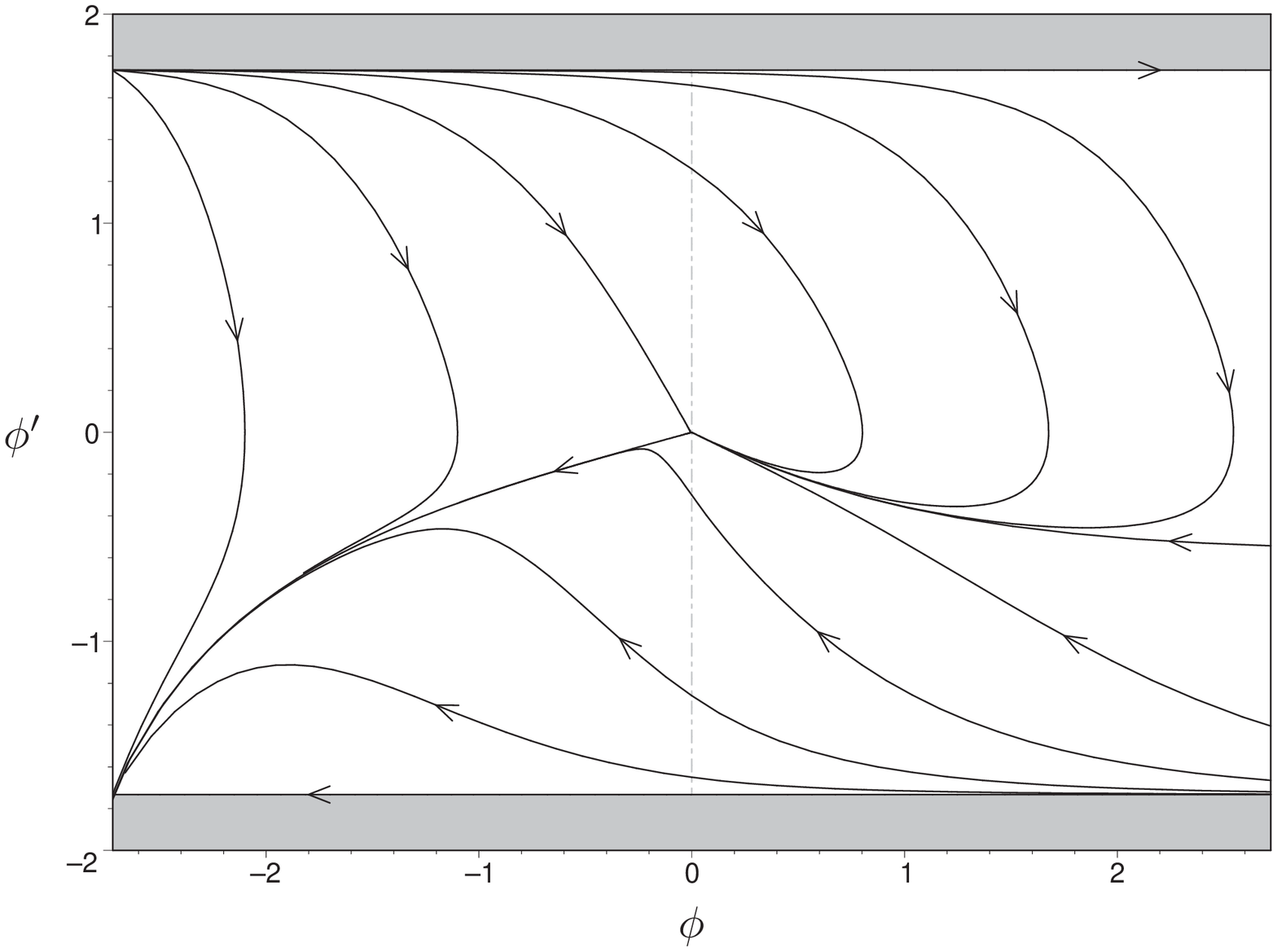}
\caption{{\sl{Example 2) 
phase portraits of the scalar field ``master'' equation 
(\ref{mejfm_abs}) in the Jordan frame (left) and its analogue (\ref{medn}),
(\ref{alpha_m}) 
in the Einstein frame (right).}}}
\end{center}
\end{figure}

The Einstein frame view 
with the canonical scalar field kinetic term 
is obtained from
Eq. (\ref{redef}), the solution is given by 
\be \label{efm_psitophi}
\pm \phi = 
\left \{ 
\begin{array}{ll}  
  \sqrt{3} \, {\rm arctanh} (\sqrt{1-\Psi}) \,, & 
\Psi \leq 1 \,, \\ 
 - \sqrt{3} \, {\rm arctan} ( \sqrt{\Psi - 1} ) \,, & \Psi \geq 1 \,,
\end{array}
\right.
\ee
see Fig. 1 right. As in the previous case, the solution has two branches
($I+,II+$) and ($I-,II-$)
related to the $\mp$ sign in Eq. (\ref{redef}) and to be interpreted as
two equivalent Einstein frame copies of the Jordan frame dynamics.
(Actually the transformation (\ref{efm_psitophi}) is infinitely 
many-valued in the domain $\Psi \geq 1$, since for each $\Psi$
we have $-\sqrt{3} \, {\rm arctan} (\sqrt{\Psi - 1}) = 
\sqrt{3}(\phi_c + n \pi)$, 
$\phi_c \in [-\frac{\pi}{2}, 0]$, but in what
follows we ignore this extra complication and assume 
$n = 0$.)

Let us focus on one of these branches by taking the $+$ sign in Eq.
(\ref{efm_psitophi}). Then $\Psi \in (0, 1]$ gets mapped onto 
$\phi \in (\infty, 0]$ and $\Psi \in [1, \infty)$ gets mapped onto
$\phi \in [0, -\frac{\pi}{2}\sqrt{3})$.
The Einstein frame field equations 
have the same form as in the example considered previously, (\ref{ef_00})--(\ref{ef_deq}),
but with the coupling function $\alpha(\phi)$ given by
\be \label{alpha_m}
\alpha(\phi)
  =  
\left \{ \begin{array}{ll}
\frac{1}{\sqrt{3}}\ {\rm tanh}\ \left(\frac{\phi}{\sqrt{3}} \right) 
& \phi \geq 0 \,,
\\  
  - \frac{1}{\sqrt{3}}\ {\rm tan}\ \left(\frac{\phi}{\sqrt{3}} \right)
& \phi \leq 0 \,. 
\end{array}
\right.
\ee
The limit to general relativity corresponds to the value $\phi = 0$
as before.

The ``master'' equation for $\phi$ retains its form (\ref{medn})
as well, but with the coupling function (\ref{alpha_m}).
The corresponding Einstein frame phase portrait on Fig. 4 right exhibits no 
symmetry reflecting the fact that we have chosen only one branch 
of $\phi(\Psi)$.
(The other branch would have given a mirror portrait with $\phi \rightarrow 
-\phi$.)
The point $(\phi=0,\phi'=0)$ is still a fixed point, but characterised
by different properties with respect to the regions 
$\phi \geq 0$ and $\phi \leq 0$. For $\phi \geq 0$ it is an attractor, 
but for $\phi \leq 0$ it is a saddle point. 

Despite the properties of this fixed point being the same 
in the respective regions of the Einstein and Jordan frame, the
phase portraits are clearly not equivalent in the two frames.
While the Jordan frame trajectories are unable to cross the general relativity
limit $\Psi=1$, the generic Einstein frame trajectories do it once.
In particular, all the Jordan frame trajectories with $\Psi<1$ 
converge to the general relativity fixed point, 
but only some of the corresponding 
Einstein frame trajectories with $\phi>0$ are collected by the fixed point 
while others pass through $\phi=0$ and get repelled from general relativity.
Similarly, all the generic Jordan frame trajectories with 
$\Psi>1$ can only get 
arbitrarily close to general relativity, 
but in the Einstein frame only some of the corresponding trajectories 
with $\phi<0$ are repelled while some 
can pass through $\phi=0$ and end up at the fixed point.
Therefore, although the issue of the Einstein
frame trajectories jumping 
from one branch to another does not arise in this case, the problem of the 
losing the correspondence between the Jordan and Einstein frame dynamics
at the general relativity limit is still manifest.


\section{Discussion}

1. {\sl General relativity as a  late time attractor 
for generic scalar-tensor theories}.
Studies of this question have usually relied on 
the Einstein frame where the equations are mathematically 
less complicated. Damour and Nordtvedt \cite{dn} 
investigated   Eq. (\ref{medn})   in the linear approximation 
of an arbitrary coupling function
at the point of general relativity ($\phi = 0$), assuming
$\alpha (\phi) \sim \phi$ which corresponds to a quadratic
``potential'' $P(\phi)  \sim \phi^2$, introduced as $\alpha \equiv dP/d\phi$. 
In the case of dust matter 
they found an oscillatory behaviour of the
scalar field with late-time relaxation to general relativity.
In comparison, Serna et al \cite{san} 
obtain $\alpha(\phi) \sim |\phi|$ 
for small values of $\phi$
from the examples of Barrow and Parsons \cite{bp} in the Jordan frame.
Now the corresponding ``potential'' has no minimum,
$P \sim {\rm sign}(\phi) \ \phi^2$, and general relativity ($\phi =0$)
is a point of inflection making possible also repulsion from
general relativity.

Both these two cases are contained in our examples
as a linear approximation near $\phi=0$: Eq. (\ref{alpha})
implies $\alpha(\phi) \sim \phi$ and Eq. (\ref{alpha_m})
implies $\alpha(\phi) \sim |\phi|$. 
The respective
qualitative behaviour can be inferred from the phase
portraits (Fig. 3, 4 right) in the neighbourhood of the fixed point 
$(\phi=0,\phi'=0)$.
Also recall that the first case involved allowing $\phi$ to pass
from one sign in Eq. (\ref{ef_psitophi}) to another, while in the second case
$\phi$ was evolving according to Eq. (\ref{efm_psitophi}) with a fixed sign.

In fact, using our phase portraits it is also possible to 
combine portraits for the cases of $\alpha(\phi) \sim -\phi$ and 
$\alpha(\phi) \sim - |\phi|$. 
Gluing together the left half of Fig. 4 right ($\phi \leq 0$) 
with its 
image under the transformation $\phi \rightarrow -\phi, \phi' \rightarrow
-\phi'$ gives the phase portrait for $\alpha(\phi) \sim -\phi$,
generically characterised by repulsion from general relativity. 
Reflection $\phi \rightarrow -\phi$ of the full Fig. 4 right yields the
portrait for $\alpha(\phi) \sim - |\phi|$ with properties similar to
the $\alpha(\phi) \sim |\phi|$ case.

It is clear that the possibility of general relativity being an
Einstein frame 
attractor crucially 
depends on the form of the coupling function $\alpha(\phi)$ and
without knowing it at least in the neighbourhood of general
relativity no conclusions can be drawn. This is in accord with the 
results of G\'erard and Mahara \cite{gm} who tried to determine
a generic behaviour around the general relativity in the Einstein frame
without specifying the coupling function and concluded that 
the ``potential'' $P$ can but need not be bounded from below. 

However, if we want to translate the results into the Jordan frame
description
the Einstein frame analysis is not reliable, as conjectured by the
general remarks in section 2 and explicitly demonstrated by the
two examples in section 3. For the Jordan frame 
conclusions about the 
STT convergence to general relativity the analysis must be carried out in the 
Jordan frame.

2. {\sl Non-minimally coupled scalar-tensor theory}. 
Sometimes a different action of scalar-tensor theory is considered 
\cite{faraoni, futamase_fakir} 
\beq \label{xi}
S_{\xi} = \frac{1}{2 \kappa^2} \int d^4 x \sqrt{-g} \left[(1 - \xi \kappa^2
\phi^2) \, R -
g^{\mu\nu} \partial_{\mu} \phi \partial_{\nu} \phi \right] + S_{matter} \,.
\eeq
It is equivalent to the action (\ref{jf4da}) of the 
scalar-tensor theory in the Jordan frame with a specific
coupling function $\omega$, if
  a redefinition of the scalar field is performed,
\beq \label{trans}
\frac{d \Psi}{d \phi}  = \mp \sqrt{\frac{\Psi}{\omega(\Psi)}} \,.
\eeq
However, analogously to the redefinition (\ref{redef}) 
it (i) contains a sign ambiguity and
(ii) is singular at the limit to general relativity, $\omega \rightarrow
\infty$.
It seems that the actions $S_{\xi}$ and $S_{_{\rm J}}$ are not
equivalent at the limit to general relativity
since $S_{_{\rm J}}$ is obtained from $S_{\xi}$ through a
singular transformation (\ref{trans}).

Note that Faraoni \cite{faraoni1} has also recently pointed out
that the correspondence between modified $f(R)$ theories and 
scalar-tensor theories of gravity breaks down in the limit to
general relativity. This indicates that general relativity
may be a rather special theory for its different modifications.

3. {\sl PPN}. We have demonstrated that there are essential
differences at the limiting process to general relativity
between the  scalar field $\Psi$ in the Jordan frame and 
the canonical scalar field $\phi$ in the Einstein frame. 
In principle, the differences may be  reflected  in  present day
observations, but only indirectly, 
through possible differences in the form of the solutions
for the scalar fields.  
The Eddington parameters which 
determine direct observational consequences 
and are given in terms of the coupling function $\omega(\Psi)$
in the Jordan frame \cite{will, nordtvedt70}
depend only on the
quantities without sign ambiguity in the Einstein frame \cite{dn},
\be
\alpha^2 (\phi) = \frac{1}{2 \omega (\Psi(\phi))+3} \,, \qquad \qquad
\frac{d\alpha}{d\phi} = 
\frac{2}{G(t_0)}\frac{(2\omega(\Psi(\phi))+4)}{(2\omega(\Psi(\phi)) +3)^3}
\frac{d\omega}{d\Psi} \,,
\ee
where $G(t_0)$ is the present day measured gravitational constant.

\section{Conclusion}

The action functionals $S_J$ and $S_E$ of the Jordan and the Einstein
frame description are equivalent in the sense that they are
connected by  conformal transformation of the metric
and redefinition of the scalar field.
However, at the limit of general relativity  
the redefinition of the scalar field is singular and the
correspondence between the different frames is lost.
This results in a different behaviour of solutions of the field
equations at this limit, e.g., in our examples of FLRW cosmology, the scalar
field $\Psi$ in the Jordan frame never crosses its general
relativistic value $\Psi_0 =1$, but scalar field $\phi$ in the
Einstein frame may oscillate around its general relativistic
value $\phi_0=0$. 
We argue that these solutions cannot
be properly set into correspondence using the redefinition of 
the scalar field (\ref{redef}). 
In order to investigate the scalar field as it  approaches to the limit of general relativity,
we must choose the frame from the very beginning by
using some additional assumptions. If our choice is that the Jordan
frame is basic, then the attractor mechanism towards
general relativity  must be reconsidered in the Jordan frame
\cite{meiemeie}.

\bigskip

{\bf Acknowledgements}

\medskip

We are grateful to the anonymous referee whose insightful comments helped us to clarify the 
presentation of Secs. I and II.
M.S. acknowledges useful comments by J.P. Mimoso and S.D. Odintsov at the Bilbao 
meeting BICOS 2007.
This work was supported by the Estonian Science Foundation
Grant No. 7185 and by Estonian Ministry for Education and Science
Support Grant No. SF0180013s07.

\end{document}